# FLEXIBLE AND HIGHLY CONDUCTIVE COMPOSITES BY IMPREGNATION OF POLYDIMETHYLSILOXANE IN GRAPHITE NANOPLATES PAPER

**Daniele Battegazzore, Alberto Fina**

Dipartimento di Scienza Applicata e Tecnologia, Politecnico di Torino, Alessandria Campus, Viale Teresa Michel 5, 15121 Alessandria, Italy

Corresponding author: alberto.fina@polito.it

## Abstract

This paper deals with the preparation and characterization of conductive and flexible sheets based on graphite nanoplates (GNP) and polydimethylsiloxane (PDMS). highly porous GNP nanopaper were prepared by filtration from a GNP suspension in solvent. Subsequently, the impregnation strategy was pursued to obtain a composite material. By varying the concentration of the solution to be impregnated and the deposition time, PDMS/GNP nanopapers are produced with a range of PDMS content, porosity and density. Thermal diffusivity of the films (both in-plane and cross-plane) were measured and correlated with the structure of the nanopapers. Based on the results found and considering the restraints, three formulations were further investigated for their physical, thermal and mechanical properties. Highly flexible, tough and conductive (up to 25 W/mK) nanopapers were obtained, for possible application in modern electronic devices.

## Materials and Methods

### *Materials*

Graphite nanoplatelets (GNP) used in this work was kindly supplied by AVANZARE (Navarrete, La Rioja, Spain) and prepared via rapid thermal expansion of overoxidized-intercalated graphite, as previously reported [1] and used as supplied without any further treatments. Dimethylformamide (DMF) (Sigma-Aldrich 99.8%) and hexane (Carlo Erba purity >95%) was purchased from Sigma-Aldrich.

### *Nanopapers production*

A suspension containing 25 mg of GNP and 75 mL of N, N-dimethylformamide (DMF) was prepared inside a 250 ml glass beaker and sonicated with immersion tip (Sonics Vibracell VCX-750, Sonics &Materials Inc. a 13 mm diameter Ti-alloy tip; Time 15 min, Pulse 30 sec ON 30 sec OFF, Power 500 W, Amplitude 30%). The suspension was then removed from the sonicator and after about 30 minutes of cooling, it was poured onto a polyamide filtering membrane of 47 mm diameter and porosity of 0.45 µm and slowly filtered with the help of a vacuum pump (10-15 minutes). First the filtered film was covered with 30 mL of ethanol and then with 30mL of diethyl ether, vacuuming each of the two solvents to eliminate them and finally leaving the pump attached for 10 minutes to ensure that the sheet was dry. Then, the GNP supported onto the membrane was dried in a vacuum oven at 70°C for at least 2h in order to eliminate DMF still present and carefully detachment to obtain self-standing nanopapers. GNP nanopapers were placed on a Teflon surface and impregnated with polydimethylsiloxane (PDMS, Sylgard® 184 by Dow) in hexane, with variable concentration, namely 9% (1:10), 17% (1:5) and 29% wt. (1:2.5). Crosslinker (Sylgard® 184 Curing Agent by Dow) was also added in the solution at 10% with respect to PDMS. A volume of 0.5 ml for each impregnation steps was homogeneously dripped manually on the film surface using a pipette and let dry under hood at room temperature for 30 minutes. Finally, the sample was placed in an oven at 70°C for 24 hours for cross-linking. After this time, the composite was carefully detached from the Teflon support. The samples have been coded in the form ***c% nx***, where c is the concentration (wt.%) of the solutions and n is the number of impregnation steps. Larger nanopapers (90mm diameter, referred to as **L**, following the above defined materials coding) were also prepared, scaling up suspension to 108 mg GNP in 327 mL DMF. To prevent excessive heating and evaporation of DMF during sonication, a larger beaker was used placed it in a water bath at a temperature of about 5°C with different operating conditions (Time 20 min, Pulse 15 sec ON 30 sec OFF, Power 500 W, Amplitude 30%). Selected nanopapers underwent compression during the polymer crosslinking (referred to as **P**, following the above defined materials coding), under a pressure of 2.2 MPa for 2h at a temperature of 70°C, using a manual hydraulic press (Specac) between two Teflon plates as supports. The crosslinking phase was found to be much slower than the not pressed samples, thus the nanopaper was left in an oven for 16h at 70°C.

*Characterization methods*

To calculate the densities, nanopapers were die cut into 25 mm diameter disks and the method of weighing the disks in air and in water according to the ASTM D792, ISO 1183 standard was performed. The density found is the one defined as $\rho_{sample}$ in the text.

The volume percentage of voids in the sample was calculated by the formula:

$$Porosity\ (vol.\%) = \left(1 - \left(\frac{\rho_{sample}}{\left(\frac{P_{GNP}}{P_{impregnated}}\right)\cdot \rho_{GNP} + \left(1 - \frac{P_{GNP}}{P_{impregnated}}\right)\cdot \rho_{PDMS}}\right)\right)\cdot 100$$

where $\rho_{GNP} = 2{,}267$ g/cm$^3$ and $\rho_{PDMS} = 0{,}958$ g/cm$^3$, P$_{GNP}$ and P$_{impregnated}$ are the nanopaper weight before and after impregnation. The same weights (P$_{GNP}$ and P$_{impregnated}$) were used to evaluate the wt.% of GNP present in the composite.

$$\text{wt.\% GNP} = \left[1 - \frac{(P_{impregnated} - P_{GNP})}{P_{impregnated}}\right] * 100$$

$$GNP(vol\%) = (100 - porosity) * \left(\frac{\frac{wt.\%\ GNP}{\rho_{grafene}}}{\frac{wt.\%\ GNP}{\rho_{grafene}} - (100 - wt.\%\ GNP) * \rho_{PDMS}}\right)$$

$$PDMS(vol\%) = 100 - GNP(vol.\%) - Porosity(vol.\%)$$

The thermal diffusivity (α) of the prepared nanopapers was measured at 25°C using a Netzsch LFA 467 Hyperflash xenon light flash analysis (LFA) on the same 25 mm diameter specimens. The nanopapers were measured for both cross-plane and in-plane diffusivity. Measurements were carried out with consistent flash parameters (230V, 200μs) and repeated five times for each specimen to get average thermal diffusivity and standard deviation.

Thermal conductivity was calculated from the measured diffusivity values, multiplied by the density and specific heat capacity of the different materials. The value of the specific heat capacity for impregnated nanopapers was calculated as a weighted average with respect to the wt.% of GNP and PDMS in the sample

$$C_{p\ sample} = C_{p\ GNP} \cdot \frac{wt.\%\ GNP}{100} + C_{p\ PDMS} \cdot \frac{(100 - wt.\%\ GNP)}{100}$$

with $C_{p\ GNP}$ = 0.71 kJ/kg * K; $C_{p\ PDMS}$ = 1.46 kJ / kg * K.

Tensile tests were conducted using the dynamometer Instron Model 5966 and a 50 N load cell. The pneumatic clamps chosen were equipped with flat 25x25mm$^2$ smooth faces. The initial distance between the two clamps was set at 20mm and a preload force at 0.02 N to ensure the sample straightness at the start of the test. A crosshead speed of 0.1 mm/min was established and the end of the test was considered when the force was below the value of 0.1 N. The tests gave data on the elastic modulus (E), the maximum stress ($\sigma_{max}$) and the elongation at maximum stress ($\varepsilon_{max\ stress}$) of the samples. The samples for the mechanical properties were obtained by cutting specimens of 10x40 mm$^2$ size, three samples were tested for each formulation.

Morphological characterization of both nanopaper and nanocomposites was performed by a high-resolution field emission scanning electron microscope (FESEM, MERLIN 4248 by ZEISS) operated at 3 kV and Scanning Electron Microscope (SEM, EVO 15 by Zeiss) at beam voltage of 20 kV and working distance of 8.5 mm. Nanopaper and composites were cryofractured in liquid nitrogen and directly observed without any further preparation.

## Results and Discussion

Pristine GNP nanopapers were observed by SEM and expectedly found highly porous (Figure 1), with thickness in the range of 200-300 μm and density in the range of 0.07-0.12 mg/mm$^3$. This structure is perfectly suitable for solution impregnation and led to an optimal absorption of PDMS solution, until 0.5 ml volume, without significant visible excess, thanks to the rapid evaporation of the solvent during deposition.

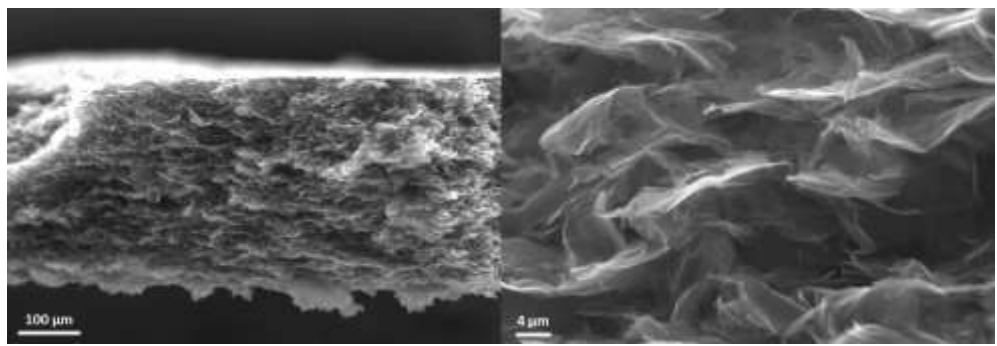

*Figure 1. SEM micrographs of pristine GNP nanopaper.*

In a first phase of this work, impregnation tests were carried out with different PDMS concentration. FESEM images of cross sections and surface morphologies of the 9% 1x (a), 17% 1x (b) and 29% 1x (c) samples are reported in (Figure 2). As expected, upon the evaporation of hexane, a highly porous structure is retained, as clearly evidenced by the presence of many air pockets sizing between a few µm to a few tens of µm.

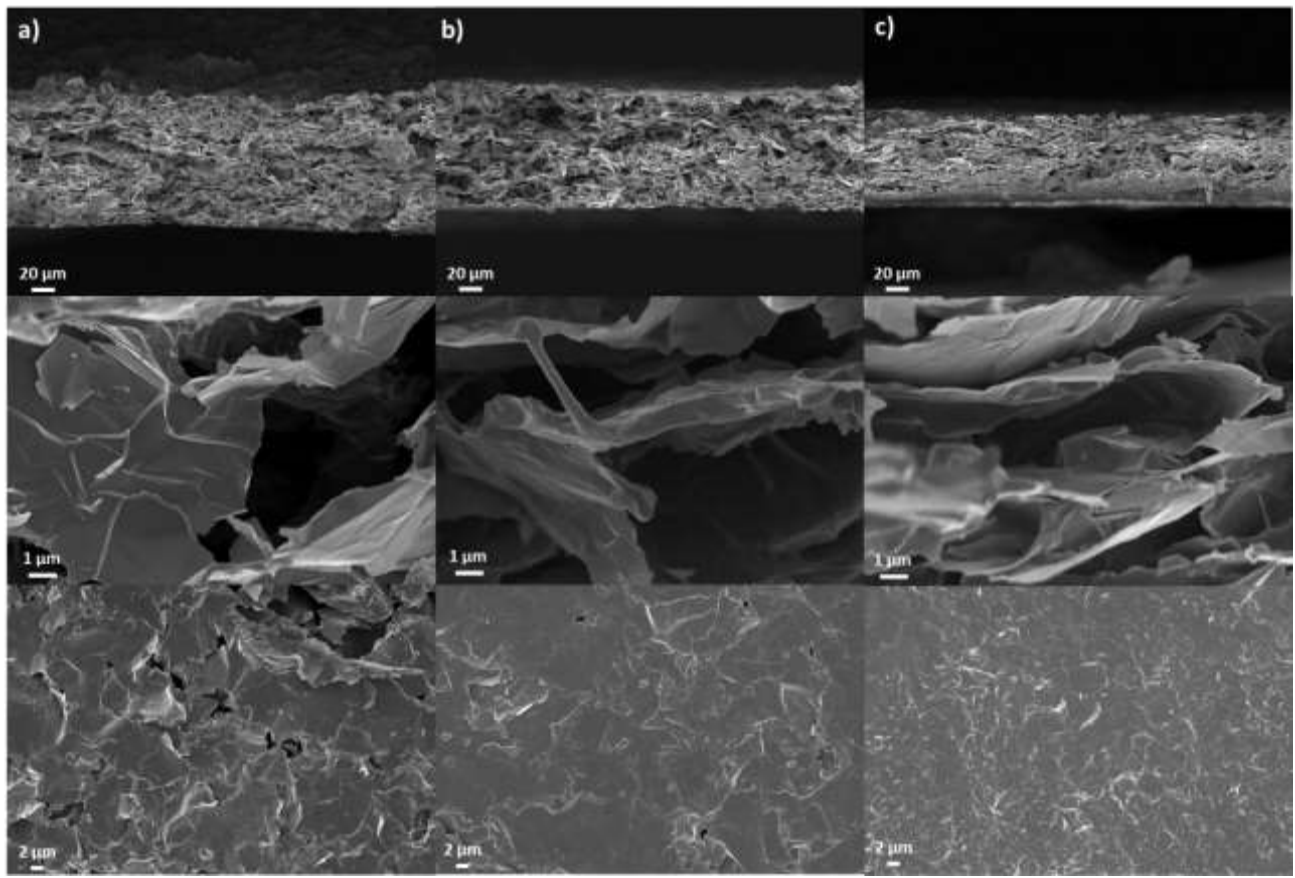

*Figure 2: FESEM micrographs of GNP nanopaper impregnated with 9% 1x (a) 17% 1x (b) and 29% 1x (c) solution.*

Aiming at the nanopaper porosity reduction, multiple impregnations of the various solutions at different concentrations are investigated (coded as 2x and 3x). Density ($\rho$) of the nanopapers was found to increases with the number of impregnations and the porosity is progressively reduced (Table 1). By this approach, low porosity conditions are obtained by filling the voids with the polymer and therefore lowering the relative percentage of GNP in the composite. Therefore, progressive densification of the nanopaper leads to a lower GNP fraction. For instance, a density of approx. 1 mg/mm$^3$ was obtained for 29% 3x nanopaper, corresponding to 10wt.% GNP content. Furthermore, it is worth noting that similar wt.% contents of GNP and vol.% porosity can be achieved with different impregnation conditions (e.g. 17% 2x vs. 29% 1x).

*Table 1: Density, porosity and thermal diffusivity values calculated for impregnated nanopapers*

| Sample code | ρ (g/cm³) | Wt.% GNP | Vol.% porosity | Thermal Diffusivity α (mm²/s) | |
|---|---|---|---|---|---|
| | | | | Cross-Plane | In-Plane |
| **9% 1x** | 0.27±0.05 | 47.8±1.0 | 83±4 | 1.44±0.05 | 51.3±2.1 |
| **9% 3x** | 0.79±0.05 | 25.9±0.1 | 39±4 | 1.24±0.05 | 24.2±0.4 |
| **17% 1x** | 0.48±0.05 | 28.7±0.1 | 64±4 | 0.93±0.05 | 30.8±0.4 |
| **17% 2x** | 0.60±0.05 | 24.7±0.1 | 54±4 | 1.26±0.03 | 14.6±0.8 |
| **29% 1x** | 0.67±0.05 | 22.5±0.1 | 47±4 | 1.33±0.01 | 12.7±0.2 |
| **29% 2x** | 0.85±0.05 | 16.4±0.1 | 28±4 | 0.75±0.01 | 12.2±0.4 |
| **29% 3x** | 1.03±0.05 | 10.3±0.1 | 7±4 | 1.00±0.01 | 8.6±0.4 |

Thermal diffusivity of the different impregnated nanopapers were evaluated by light flash analyses, in both in-plane and cross-plane modes (Table 1), for correlation with nanopaper density and GNP content (Figure 3).

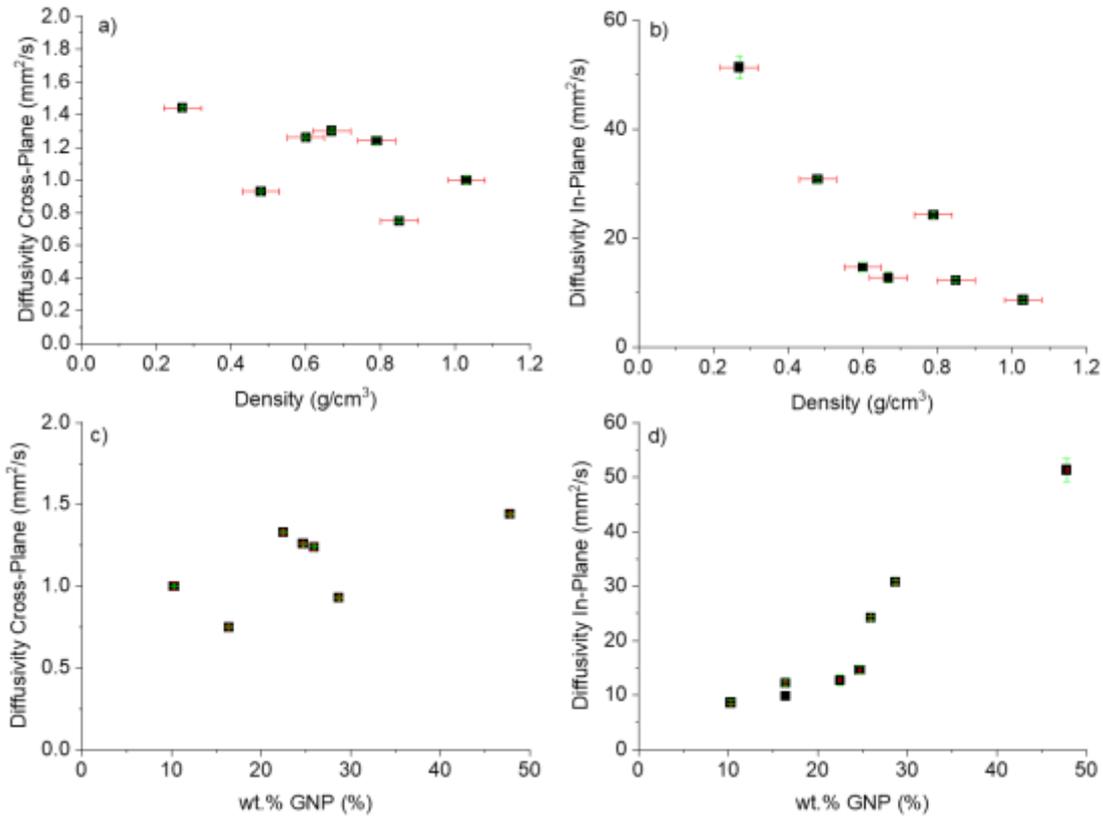

*Figure 3: Thermal diffusivity of PDMS impregnated nanopapers, as a function of nanopaper density (a,b) and as a function of GNP content (c.d)*

PDMS/GNP nanopaper expectedly lower than pristine GNP nanopaper (cross plane 9.2±0.1 mm²/s, in plane 101±5 mm²/s). By analyzing the diffusivity data as a function of density *Figure* 3 a b, it is possible to verify a decreasing trend for both the cross-plane and in-plane diffusivity, with increasing density. This fact is explained by the increased quantity of PDMS that stands between the GNP lamellae. Being PDMS a poorly conductive material, the increase in PDMS content is reflected in a decrease in the diffusivity in the composite, which is more evident in the in-plane orientation. Conversely, thermal diffusivity show an expected increase as a function of GNP content (*Figure 3c,d*), which is particularly remarkable for in-plane measurements. Based on these preliminary impregnation tests, selected formulations were further investigated preparing larger (L) nanopapers, targeting GNP content of ~50wt.%GNP (nanopaper 9% 1x L), ~25% wt.%GNP (nanopaper 9% 3x L) and ~15% wt.%GNP (nanopaper 29% 2x L). The replicated formulations were characterized for the densities and thermal properties and the data are reported in Table 2.

*Table 2: Density, porosity and thermal diffusivity values calculated for GNP and PDMS/GNP (L) nanopapers*

| Sample code | $\rho$ (g/cm$^3$) | wt.% GNP | Vol.% porosity | Diffusivity $\alpha$ (mm$^2$/s) | | Conductivity (W/m*K) | |
|---|---|---|---|---|---|---|---|
| | | | | Cross-Plane | In-Plane | Cross-Plane | In-Plane |
| **GNP** | 0.12±0.01 | 100 | 95±2 | 9.25±0.06 | 101.2±4.8 | 0.79±0.06 | 8.6±0.4 |
| **GNP-P** | 0.37±0.02 | 100 | 83±2 | 2.74±0.03 | 152.3±2.2 | 0.71±0.06 | 39.5±0.6 |
| **9% 1x L** | 0.11±0.02 | 55.6±0.1 | 93±4 | 1.93±0.01 | 44.5±1.8 | 0.23±0.06 | 5.4±0.2 |
| **9% 3x L** | 0.28±0.02 | 30.3±0.1 | 79±4 | 2.91±0.01 | 34.4±1.9 | 1.00±0.06 | 11.9±0.7 |
| **9% 3x L-P** | 0.91±0.03 | 31.4±0.6 | 27.3±2.2 | 2.11±0.88 | 22.4±0.6 | 2.38±0.96 | 25.4±0.2 |
| **29% 2x L** | 0.60±0.02 | 13.2±0.1 | 47±4 | 1.02±0.01 | 7.8±0.3 | 0.83±0.06 | 6.4±0.3 |

For comparison, pristine GNP papers were also prepared and analyzed for their thermal conduction properties, both as obtained from filtration of GNP suspension and after compression to consolidate their highly porous structure. Results reported in Table 2 confirmed a significant density increase upon compression, still associated to a very high porosity in the range of 80%, which was also confirmed by SEM imaging (Figure 4). Indeed, a significant reduction in thickness from about 300 to 80 μm was observed upon compression of GNP nanopaper, along with a higher in-plane orientation of nanoflakes. These differences in the nanopaper microstructure are also reflected in the thermal diffusivity values for GNP nanopapers. Indeed, the better horizontal orientation of GNP platelets allows to have an increase of in-plane diffusivity while on the contrary decreases the cross plane diffusivity. Thanks to the densification of the sample, the cross plane conductivity turns out to be practically equal between the pristine and pressed sample while in in-plane direction it has more than quadrupled (Table 2) to about 40 W/mK. It is worth mentioning that this value is relatively low compared to previously reported GNP nanopapers [2-4], owing to its low density. Beside the thermal conductivity performance, it has to be mentioned that pristine GNP nanopapers are intrinsically brittle, which limits their practical application as heat spreaders.

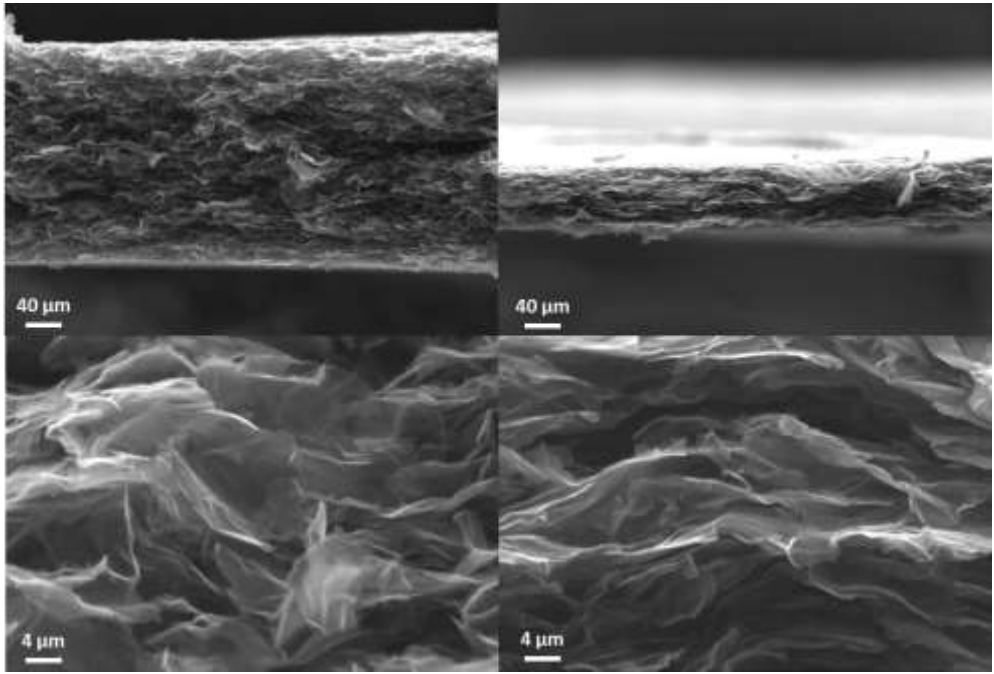

*Figure 4. SEM micrographs for GNP and GNP-P nanopapers*

The heat conduction properties in cross-plane and in-plane for standard PDMS/GNP are also summarized in Figure 5, showing best performance is obtained with triple deposition of PDMS 9% solution. Indeed, while diffusivity values are directly dependent on the GNP fraction, the increase in density obtained when progressively filling voids with PDMS is also affecting the thermal conductivity of the nanopaper, leading to best performance for the intermediate GNP concentration.

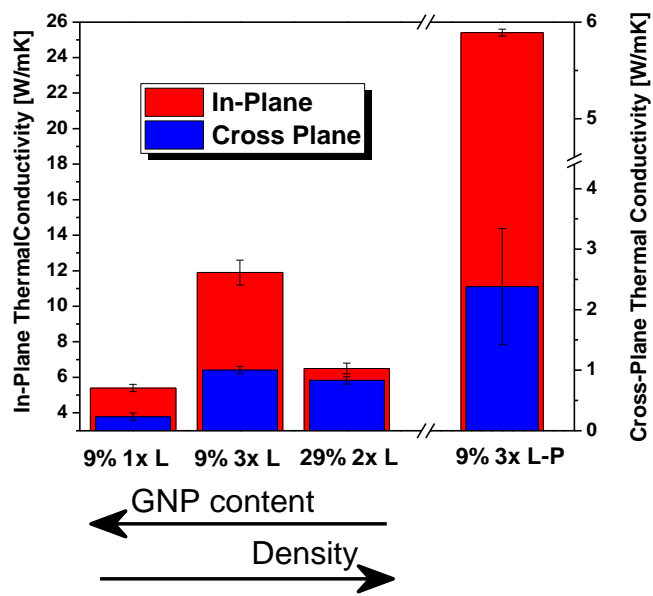

*Figure 5: In plane anc cross-plane thermal conductivities for PDMS/GNP nanopapers*

To further enhance nanopaper densities, a 9% 3x L impregnated paper underwent mechanical compression during crosslinking, resulting in a 3-fold increase in density, leading to a dramatic enhancement in in plane thermal conductivity to 25 W/mK. SEM micrographs comparison between standard vs pressed nanopapers (Figure 6) confirmed the reduction in the thickness for compressed PDMS/GNP nanopaper, as well as a denser structure, with more evident polymer flaps connecting the various GNP lamellae.

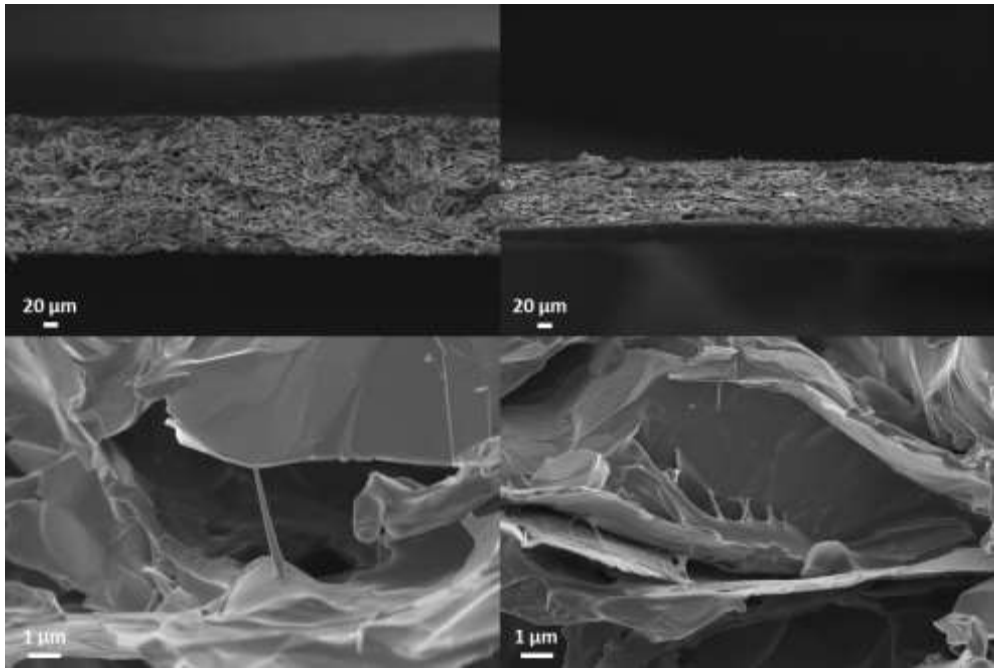

*Figure 6. SEM micrographs for PDMS/GNP nanopapers 9% 3x L and 9% 3x L-P*

Envisaging application of PDMS/GNP nanopapers as flexible heat spreader, mechanical properties were investigated by tensile test. The elastic modulus (E), the maximum stress ($\sigma_{max}$) and the elongation at maximum stress ($\varepsilon_{max\,stress}$) were evaluated and reported in Table 3 and Figure 7.

*Table 3: Mechanical properties of GNP and PDMS/GNP nanopapers*

| Sample | E (MPa) | $\sigma_{max}$ (MPa) | $\varepsilon_{max\,stress}$ (%) |
|---|---|---|---|
| PDMS | 1.7±0.2 | 3.98±0.40 | 135.5±3.3 |
| GNP | 31.1±3.1 | 0.13±0.03 | 1.1±0.2 |
| 9% 1x L | 22.6±1.6 | 0.17±0.01 | 1.3±0.2 |
| 9% 3x L | 52.0±13.9 | 0.45±0.01 | 1.7±0.3 |
| 9% 3x L-P | 114 | 0.78 | 1.25 |
| 29% 2x L | 49.4±7.9 | 0.44±0.08 | 2.2±0.5 |

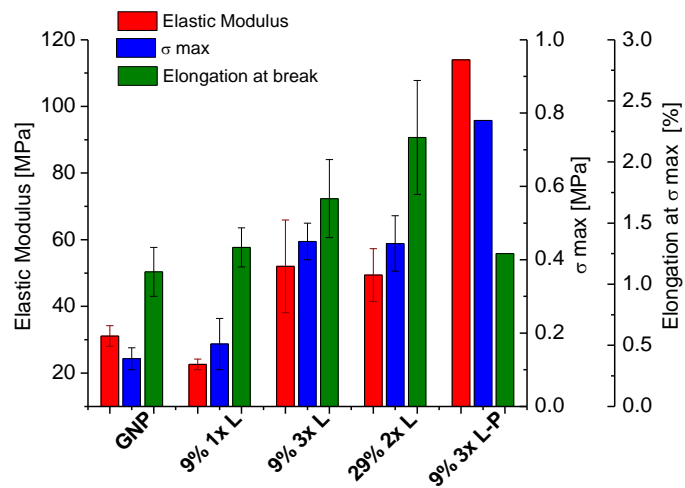

*Figure 7: Mechanical properties of PDMS/GNP nanopapers*

Pristine GNP nanopapers exhibit limited mechanical properties, with elastic modulus of 31 MPa, 0.13MPa resistance and about 1% deformation at break, suggesting mechanical performance to be controlled by the weak interactions between graphite nanoplates. The addition of soft and deformable PDMS, acting as a binder in the GNP network, generally enhanced mechanical properties in all regards of stiffness, resistance and elongation at break. Best performances in terms of both elastic modulus and max resistance are confirmed for 9% 3x L-P nanopaper.

## Conclusions

Impregnation of PDMS within the porous structure of a nanopapers based on graphite nanoplates was successfully carried out by a simple casting method from a PDMS solution in hexane. The obtained PDMS/GNP nanopapers porosity, obtained as an effect of the solvent evaporation, can be adjusted depending on the concentration of the PDMS solution and the number of impregnation steps. Thermal diffusion of the nanopapers was correlated with both the density of the nanopapers and the GNP contents, leading to in-plane diffusivity values in the range between 10 and 50 mm$^2$/s. Depending on the nanopaper density and composition, the nanopapers thermal conductivity (in-plane) was obtained in the range between 5 and 25 W/mK. PDMS crosslinking under pressure to densify the nanopapers was proven to enhance both thermal conductivity and mechanical properties, with dramatic reinforcing (approx. 4x in elastic modulus, 6x in max resistance) compared to pristine GNP paper. Based on the results obtained, PDMS/GNP nanopapers appear to be promising solution in heat spreader application requiring flexibility and toughness, such as in modern flexible electronics, as well as in wearable and implantable devices.

## Acknowledgments

This work has received funding from the European Research Council (ERC) under the European Union's Horizon 2020 research and innovation programme grant agreement 639495 — INTHERM — ERC-2014-STG. Julio Gomez at Avanzare Innovación Tecnólogica € is gratefully acknowledged for providing GNP. Erica Fadda is acknowledged for her experimental work. Mauro Raimondo and Dario Pezzini at Politecnico di Torino are also acknowledged for SEM observations.